\documentclass[aps,prl,twocolumn,showpacs,groupedaddress]{revtex4}
\usepackage{graphicx}
\usepackage{dcolumn}
\usepackage{bm}
\usepackage{amssymb}
\usepackage{psfrag}%
\usepackage{amsmath}%
\usepackage{amsfonts}	
\providecommand{\U}[1]{\protect\rule{.1in}{.1in}}
\def\met{\mbox{${\hbox{$E$\kern-0.6em\lower-.1ex\hbox{/}}}_T$ }}
\begin{document}

\title{Measurement of the ratio of the $p\bar{p}\rightarrow W$+$c$-jet
cross section to the inclusive $p\bar{p}\rightarrow W$+jets cross section}
%
\author{V.M.~Abazov$^{36}$}
\author{B.~Abbott$^{75}$}
\author{M.~Abolins$^{65}$}
\author{B.S.~Acharya$^{29}$}
\author{M.~Adams$^{51}$}
\author{T.~Adams$^{49}$}
\author{E.~Aguilo$^{6}$}
\author{S.H.~Ahn$^{31}$}
\author{M.~Ahsan$^{59}$}
\author{G.D.~Alexeev$^{36}$}
% ...

\affiliation{\vspace{0.1 in}(The D\O\ Collaboration)\vspace{0.1 in}}
% ...
\affiliation{$^{6}$University of Alberta, Edmonton, Alberta, Canada,
                Simon Fraser University, Burnaby, British Columbia, Canada,
                York University, Toronto, Ontario, Canada, and
                McGill University, Montreal, Quebec, Canada}
% ...
\affiliation{$^{29}$Tata Institute of Fundamental Research, Mumbai, India}
% ...
\affiliation{$^{31}$Korea Detector Laboratory, Korea University, Seoul, Korea}
% ...
\affiliation{$^{36}$Joint Institute for Nuclear Research, Dubna, Russia}
% ...
\affiliation{$^{49}$Florida State University, Tallahassee, Florida 32306, USA}
% ...
\affiliation{$^{51}$University of Illinois at Chicago,
                Chicago, Illinois 60607, USA}
% ...
\affiliation{$^{59}$Kansas State University, Manhattan, Kansas 66506, USA}
% ...
\affiliation{$^{65}$Michigan State University,
                East Lansing, Michigan 48824, USA}
% ...
\affiliation{$^{75}$University of Oklahoma, Norman, Oklahoma 73019, USA}
% ...

\hspace{5.2in} \mbox{Fermilab-Pub-08/062-E}
\date{March 14, 2008}
\begin{abstract}
We present a measurement of the fraction of inclusive $W$+jets events
produced with net charm quantum number $\pm1$, denoted $W$+$c$-jet, in
$p\bar{p}$ collisions at $\sqrt{s}=1.96$ TeV using approximately $1$~fb$^{-1}$  of data collected by the D0 detector at the Fermilab Tevatron Collider. \ We identify the $W$+jets events via the leptonic $W$ boson decays. \ Candidate
$W$+$c$-jet events are selected by requiring a jet containing a muon in
association with a reconstructed $W$ boson and exploiting the charge
correlation between this muon and $W$ boson decay lepton to perform a nearly
model-independent background subtraction. \ We measure the fraction of $W$+$c$-jet events in the inclusive $W$+jets sample for jet $p_{T}>20$ GeV and pseudorapidity $|\eta|<2.5$ to be $0.074$$\pm0.019$(stat.)$\pm^{0.012}_{0.014}$(syst.), in agreement with theoretical predictions. \ The probability that background fluctuations could produce the observed fraction of $W$+$c$-jet events is estimated to be $2.5\times 10^{-4}$, which corresponds to a $3.5$ $\sigma$ statistical significance.
\end{abstract}
\maketitle

In hadron-hadron collisions, the $W/Z$+$b$- or $c$-jet final state can signal
the presence of new physics; however, only a few measurements
\cite{cdfZbpaper,d0Zbpaper,cdfWcpaper} of cross sections for these standard model processes exist. \ Charm quark production in association with a $W$ boson can be a significant background, for example, to top quark pair, single top quark and Higgs boson productions, and to supersymmetric top quark (stop) pair production when only the $\tilde{t}\rightarrow c \tilde{\chi}_{1}^{0}$ decay channel is allowed by the mass difference between the stop quark and the neutralino. \ Moreover, as the squared Cabibbo-Kobayashi-Maskawa matrix element, $|V_{cd}|^{2}$, suppresses the expected leading order $d$ quark-gluon fusion production mechanism, $W$+$c$-quark production provides direct sensitivity to the proton's $s$ quark parton distribution function (PDF), $s(x,Q^{2})$, where $x$ is the momentum fraction of the proton carried by the $s$-quark and $\sqrt{Q^{2}}$ is the hard scatter scale~\cite{Wctheorypaper}. \ This PDF has been measured directly only in fixed target neutrino-nucleon deep inelastic scattering experiments using relatively low momentum transfer squared, $Q^{2}$, of the order $1-100$ GeV$^{2}$~\cite{mason,d0altonMGon,strangeseapaper1,strangeseapaper2,charmII,CDHS1,CDHS2}. \ A probe of the $s$ quark PDF at the Tevatron tests the universality of $s(x,Q^{2})$ and its QCD evolution up to $Q^{2}=10^{4}$ GeV$^{2}$. \ The strange quark PDF initiates both standard model (\text{e.g.}, $sg\rightarrow W^{-}$+$c$) and possible new physics processes (\text{e.g.}, $s\bar{c}$ $\rightarrow H^{-}$)~\cite{recentCTEQpaper} at both the Fermilab Tevatron and CERN LHC colliders. \newline\indent In this Letter, we present a measurement of the cross section ratio $\sigma(p\bar{p}\rightarrow W$+$c$-jet$)$ $/\sigma
(p\bar{p}\rightarrow W$+jets$)$ as a function of jet transverse momentum $p_T$, where $W$+$c$-jet denotes a $W$ boson plus
jets final state in which the jets have a net charm quantum number $C=\pm1$,
and $W$+jets denotes any $W$ boson final state with at least one jet.
\ Several experimental uncertainties (e.g., luminosity, jet energy scale, and
reconstruction efficiency) and theoretical uncertainties (e.g.,
renormalization and factorization scales) largely cancel in this ratio. 
\newline\indent This measurement utilizes approximately $1$ fb$^{-1}$ of $p\bar{p}$ collision data at a center-of-mass energy $\sqrt{s}=1.96$ TeV collected with the D0 detector at the Fermilab Tevatron collider. \ We identify $W$ bosons through their leptonic decays, $W\rightarrow\ell\nu$, where $\ell=e,\mu$. \ $W$ bosons decaying to tau leptons are included for leptonic tau decays $\tau \rightarrow e \bar{\nu}_{e} \nu_{\tau}$ or $\tau \rightarrow \mu \bar{\nu}_{\mu} \nu_{\tau}$. \ The electron or muon from $W$ boson decays are required to be isolated, and their transverse momentum $p_{T}$ must satisfy $p_{T}>20$ GeV. \ The presence of a neutrino is inferred from the requirement that the missing transverse energy
\mbox{${\hbox{$E$\kern-0.6em\lower-.1ex\hbox{/}}}_T$ } satisfies
\mbox{${\hbox{$E$\kern-0.6em\lower-.1ex\hbox{/}}}_T$ }$>20$ GeV. \ Jets are defined using the iterative seed-based midpoint cone algorithm~\cite{d0jets} with cone radius of 0.5. \ We restrict the transverse momentum of the jet to $p_{T}>20$ GeV after it is calibrated for the calorimeter jet energy scale (JES), and its pseudorapidity to $|\eta|<2.5$, where $\eta = -\ln\left[\tan\left(\theta/2\right)\right]$ and $\theta$ is the polar angle with respect to the proton beam direction. \ We correct the jet measurement to the particle level~\cite{parjet} for comparison with theory. 
\newline\indent A muon reconstructed within a jet (a \textquotedblleft jet-muon\textquotedblright) identifies that jet (a \textquotedblleft$\mu$-tagged jet\textquotedblright) as a charm quark candidate. \ Events containing a jet-muon enrich a sample in $b$/$c$ semileptonic decays. \ Events with the jet-muon's charge opposite to or equal to that of the $W$ boson are denoted as \textquotedblleft OS\textquotedblright\ or \textquotedblleft SS\textquotedblright\ events, respectively. \ In the $W$+$c$-jet process, the charm quark decays into a muon carrying an opposite-sign charge compared to that carried by the $W$ boson, and the numbers of OS and SS events, $N_{\text{OS}}$ and $N_{\text{SS}}$, respectively, satisfy $N_{\text{OS}}\gg N_{\text{SS}}$. \ In the $W$+$c$-jet sample, $N_{\text{SS}}$ can be non-zero because a jet initiated by a $c$ quark has a small probability of containing a muon from the decay of particles other than the leading charm quark.
\ Other vector boson+jets physics processes ($W$+$g$, $W$+$c\bar{c}$, $W$+$b\bar{b}$, $Z$+jets) can produce $\mu$-tagged jets, but the charge of the jet-muon is uncorrelated with that of the boson, hence $N_{\text{OS}}\approx N_{\text{SS}}$ for these sources. \ Processes with light-quark ($u$, $d$ or $s$) initiated jets recoiling against the $W$ boson can produce a small fraction of charge-correlated $\mu$-tagged jets owing to leading particle effects~\cite{lpeffect}. \ Background from $WW$ production contributes only a small amount to the signal sample. \ The $WZ$ and $ZZ$ processes only rarely produce charge-correlated jets. \ Other final states that can produce charge-correlated jets ($t\bar{t}$, $t\bar{b}$, $W$+$b\bar{c}$ and $W$+$b$) are suppressed by small production cross sections or tiny CKM matrix elements. \ These considerations allow a measurement of the $W$+$c$-jet production rate from OS events with the backgrounds determined \textit{in situ} from SS events, up to small weakly model-dependent theory corrections. 
\newline\indent The D0 detector~\cite{d0det} is a multi-purpose device built to investigate $p\bar{p}$ collisions. \ The innermost silicon microstrip detectors followed by the scintillating fiber tracking detector, covering pseudorapidity $|\eta| \lesssim 3.0$ and located inside the 2 T superconducting solenoid, are used for tracking and vertexing. \ The liquid-argon and uranium calorimeter, a finely segmented detector in the transverse and the longitudinal directions, is used as the primary tool to reconstruct electrons, jets, and the missing transverse energy. \ It is housed in three cryostats, one central calorimeter in the region $|\eta|<1.1$ and two end caps extending the coverage to $|\eta|\approx4.0$. \ The outermost subsystem of the D0 detector is the muon spectrometer, consisting of three layers of muon tracking subdetectors and scintillation trigger counters, which is used to construct muons up to $|\eta|\approx2.0$. \ The first layer is situated before the 1.8 T iron toroid and the other two layers are outside, enclosing the detector. 
\newline\indent Candidate events in the electron (muon) decay channel of the $W$ boson must pass at least one of the single electron (muon) three-level (L1, L2 and L3) triggers used in each data taking period. \ Each level of trigger imposes tighter restrictions on the events compared to those of the preceding level. \ The single muon triggers at L1 impose hit requirements in the muon scintillators. \ Some of the triggers also require spatially matched hits in the muon tracking detectors. \ The conditions at L2 require a reconstructed muon with $p_T$ above a threshold in the range $0$ -- $5$ GeV for various triggers. \ At L3, certain triggers require a track reconstructed in the inner tracking system with $p_T>10$ GeV. \ The ratio measurement benefits from full cancellation of the trigger efficiency in the electron channel. \ This cancellation is partial in the muon channel due to the presence of two muons in the $W$+$c$-jet sample. 
\newline\indent Selection of $W\rightarrow e\nu$ candidates begins with the requirement that a cluster of energy be found that is consistent with the presence of an electron in the calorimeter. \ The cluster must: have at least $90\%$ of its energy contained in the electromagnetic part of the calorimeter; have a reconstructed track from the inner tracking system pointing to it; be isolated from other clusters in that the fraction of the energy deposited in an annulus ($0.2<$$\Delta$$\mathcal{R}=\sqrt{(\Delta\phi)^2+(\Delta\eta)^2}$ $<0.4$, where $\phi$ is the azimuthal angle) around the EM cluster is less than $15\%$ of the electromagnetic energy within the cone of radius $\Delta$$\mathcal{R}=0.2$; have longitudinal and transverse energy deposition profiles consistent with those expected for an electron; and satisfy a likelihood discriminant selection that combines tracker and calorimeter information using the expected distributions for electrons and jet background. \ The electron track's point of closest approach to the $z$-axis must be within $3$ cm of the $p\bar{p}$ interaction point, which must lie within $60$ cm of the nominal detector center. 
\newline\indent Selection of $W\rightarrow\mu\nu$ candidates begins by requiring that a muon candidate be found in the muon spectrometer with a track matched to one found in the central tracker. \ Rejection of cosmic ray background events demands that the central tracker track pass within $0.02$ or $0.2$ cm of the beam crossing point in the transverse plane, depending on whether the track is reconstructed with or without hits, respectively, in the silicon detector, and that the point of closest approach of the track should be within $3$ cm of the interaction point along the $z$-axis. \ Further cosmic ray rejection comes from scintillator timing information in the muon spectrometer. \ Requiring the $W$ boson candidate muon track to be separated from the axis formed by any jet found in the event by $\Delta$$\mathcal{R}$$(\mu,\text{jet})>0.5$ suppresses backgrounds from semileptonic decays of heavy flavor quarks in multi-jet events. 
\newline\indent For the final selection in both channels, each event must satisfy the transverse mass requirement $40\leq M_{T}\leq120$ GeV, where $M_{T}=\sqrt{2\mbox{${\hbox{$E$\kern-0.6em\lower-.1ex\hbox{/}}}_T$} p_{T}^{\ell}\left[  1-\cos\Delta\phi(\mbox{${\hbox{$E$\kern-0.6em\lower-.1ex\hbox{/}}}_T$}, p_{T}^{\ell})\right]}$ is computed from the isolated lepton $p_{T}^{\ell}$ and the $\mbox{${\hbox{$E$\kern-0.6em\lower-.1ex\hbox{/}}}_T$}$, have an azimuthal angular separation between the isolated lepton and $\mbox{${\hbox{$E$\kern-0.6em\lower-.1ex\hbox{/}}}_T$}$ directions greater than 0.4. \ Events must contain at least one jet with $p_{T}>20$ GeV after the calorimeter JES correction is applied, and $|\eta|<2.5$. \ Upon application of all selection criteria, $N_{Wj}^{e}=82747$ \ and $N_{Wj}^{\mu}=57944$ $W$+jets candidates remain in the electron and muon channels, respectively. \newline\indent Backgrounds originate from photons and jets that are misidentified as electrons and from $c\bar{c}$ and $b\bar{b}$ multi-jet events that produce an isolated muon. \ These multi-jet backgrounds are determined directly from the data using a \textquotedblleft matrix method\textquotedblright\ consisting of the following steps: \ first, \textquotedblleft loose\textquotedblright\ $W(\rightarrow\ell\nu)$+jets datasets are selected through application of all previously described selection criteria in each channel with the exception that the lepton isolation requirements are relaxed. \ This produces a set of loose candidate events, $N_{L}^{\ell}$, in each lepton channel consisting of a mixture of real $W$+jets events, $N_{W}^{\ell}$, and multi-jet background events, $N_{\text{MJ}}^{\ell}$, with $N_{L}^{\ell}=N_{W}^{\ell}+$ $N_{\text{MJ}}^{\ell}$. \ Application of the stricter lepton isolation criteria used to extract the signal changes this mixture to $N_{T}^{\ell}=\epsilon_{W}^{\ell}N_{W}^{\ell}+$ $\epsilon_{\text{MJ}}^{\ell}N_{\text{MJ}}^{\ell}$, where $N_{T}^{\ell}$ signifies the number of events in each channel satisfying the tighter isolation criteria and $\epsilon_{W}^{\ell}$ and $\epsilon_{\text{MJ}}^{\ell}$ denote the relative probabilities for loosely selected $W$ boson and multi-jet events, respectively, to satisfy the stricter isolation criteria. \ A large sample of two-jet events is used to measure $\epsilon_{\text{MJ}}^{e}$, and $\epsilon_{\text{MJ}}^{\mu}$ is estimated from a similar two-jet dataset using a sample of back-to-back muon-plus-jet events with low $\mbox{${\hbox{$E$\kern-0.6em\lower-.1ex\hbox{/}}}_T$}$. \ The factors $\epsilon_{W}^{\ell}$ are obtained from a large data sample of $Z\rightarrow\ell^{+}\ell^{-}$ events. \ Solving the equations for $N_{W}^{\ell}$ and $N_{\text{MJ}}^{\ell}$ yields estimates for the fractional contributions of multi-jet background to the inclusive $W$+jets signal of $f_{\text{MJ}}^{e}=(3.2\pm0.8)\%$ in the electron channel and $f_{\text{MJ}}^{\mu}=(4.1\pm3.0)\%$ in the muon channel. \newline\indent The channel $Z(\rightarrow\ell^{+}\ell^{-})+$jets contributes as background when one of the leptons from the $Z$ boson decay fails to be reconstructed. \ An estimate of this background follows from MC simulations of $Z$+jets production produced with \textsc{alpgen v2.05}~\cite{alpgen} using the \textsc{cteq6l}~\cite{cteq} PDF set, the \textsc{pythia v6.323}~\cite{pythia} generator for the parton fragmentation and hadronization, the \textsc{mlm} prescription~\cite{mlm} to avoid an over-counting of final state jets, and \textsc{evtgen}~\cite{evtgen} to decay the heavy hadrons. \ A \textsc{geant}~\cite{geant} based program simulates the effects of detector response, and the same reconstruction software is employed as used for data. \ This procedure results in estimates for the fractional contaminations of $f_{Z}^{e}=(0.9\pm0.1)\%$ for $Z(\rightarrow e^{+} e^{-})+$jets and $f_{Z}^{\mu}=(5.0\pm0.7)\%$ for $Z(\rightarrow\mu^{+}\mu^{-})+$jets. \ Quoted uncertainties derive mainly from systematic effects in the $Z$+jets \textsc{alpgen} cross section model that are estimated by varying the relative cross section of $W$+jets with respect to $Z$+jets by its uncertainties. \\
\begin{table*}[hptb]
\caption{Summary of quantities to estimate the $W$+$c$-jet cross section ratio. \ The first uncertainties quoted are statistical and the second systematic.}
\label{tab:eventY}
\centering
\begin{ruledtabular}
\begin{tabular}[c]{ccccc}
jet $p_T$ [GeV] & ($20$--$30$) & ($30$--$45$) & ($45$--$200$) & ($20$--$200$)  \\
\hline
\multicolumn{5}{c}{$W \rightarrow e \nu$ decay channel} \\
\hline
$N_{Wj}^{e}$ & 35695 & 24412 & 22640 & 82747  \\
$N_{\text{OS}}^{e}$  & 83 & 77 & 85 & 245  \\
$N_{\text{SS}}^{e}$  &  45 & 41 & 68 & 154  \\
$N_{\text{OS}}^{e,\text{MJ}}$  & 4.5$\pm$1.0$\pm$1.2 & 4.2$\pm$0.9$\pm$1.1 & 4.6$\pm$1.0$\pm$1.2 & 13.3$\pm$2.6$\pm$3.4 \\
$N_{\text{SS}}^{e,\text{MJ}}$  & 5.6$\pm$1.1$\pm$1.4 & 5.1$\pm$1.0$\pm$1.3 & 8.5$\pm$1.5$\pm$2.2 & 19.3$\pm$2.9$\pm$4.9  \\
$N_{\text{OS}}^{e,WW}$  & 1.8$\pm$0.6 & 2.1$\pm$0.7 & 2.3$\pm$0.8 & 6.2$\pm$2.1 \\
$N_{\text{SS}}^{e,WW}$  & 0.4$\pm$0.1 & 0.6$\pm$0.2 & 0.9$\pm$0.3 & 1.9$\pm$0.5  \\
$N_{\text{OS}}^{e,t\bar{t}}$  & 2.4$\pm$0.6 & 4.6$\pm$1.1 & 11.8$\pm$2.8 & 18.8$\pm$4.5 \\
$N_{\text{SS}}^{e,t\bar{t}}$  & 2.1$\pm$0.5 & 4.1$\pm$1.0 & 10.0$\pm$2.4 & 16.1$\pm$3.9  \\
$N_{\text{OS}}^{e,t\bar{b}}$  & 1.1$\pm$0.3 & 2.1$\pm$0.6 & 3.1$\pm$0.9 & 6.3$\pm$1.8 \\
$N_{\text{SS}}^{e,t\bar{b}}$  & 0.8$\pm$0.2 & 1.4$\pm$0.4 & 2.5$\pm$0.7 & 4.6$\pm$1.3  \\
$f_c^{e}$ & 1.183$\pm$0.017$\pm$0.018 & 1.164$\pm$0.019$\pm$0.017 & 1.118$\pm$0.024$\pm$0.017 & 1.149$\pm$0.007$\pm$0.017 \\
$\epsilon_{c}^{e}$ & 0.0113$\pm$0.0015$^{+0.0017}_{-0.0017}$ & 0.0125$\pm$0.0011$^{+0.0019}_{-0.0019}$ & 0.0125$\pm$0.0020$^{+0.0019}_{-0.0019}$ & 0.0124$\pm$0.0012$^{+0.0019}_{-0.0019}$ \\
$\frac{\sigma[W(\rightarrow e \nu)+c\text{-jet}]}{\sigma[W(\rightarrow e \nu)+\text{jets}]}$ & 0.079$\pm$0.031$^{+0.013}_{-0.022}$ & 0.100$\pm$0.038$^{+0.017}_{-0.016}$ & 0.043$\pm$0.049$^{+0.007}_{-0.007}$ & 0.073$\pm$0.023$^{+0.012}_{-0.014}$ \\
\hline
\multicolumn{5}{c}{$W \rightarrow \mu \nu$ decay channel} \\
\hline
$N_{Wj}^{\mu}$ & 27378 & 17325 & 13241 & 57944 \\
$N_{\text{OS}}^{\mu}$  & 76 & 64 & 63 & 203 \\
$N_{\text{SS}}^{\mu}$  &  28 & 38 & 56 & 122 \\
$N_{\text{OS}}^{\mu,\text{MJ}}$  & 4.6$\pm$1.8$\pm$3.3 & 3.8$\pm$1.5$\pm$2.7 & 3.8$\pm$1.5$\pm$2.7 & 12.2$\pm$4.6$\pm$8.7 \\
$N_{\text{SS}}^{\mu,\text{MJ}}$  & 2.0$\pm$1.3$\pm$1.4 & 2.8$\pm$1.7$\pm$2.0 & 4.1$\pm$2.5$\pm$2.9 & 8.8$\pm$5.4$\pm$6.3 \\
$N_{\text{OS}}^{\mu,WW}$  & 0.8$\pm$0.3 & 1.6$\pm$0.5 & 1.8$\pm$0.6 & 4.2$\pm$1.6 \\
$N_{\text{SS}}^{\mu,WW}$  & 0.3$\pm$0.1 & 0.3$\pm$0.1 & 0.6$\pm$0.2 & 1.2$\pm$0.4 \\
$N_{\text{OS}}^{\mu,t\bar{t}}$  & 1.2$\pm$0.3 & 2.3$\pm$0.6 & 5.8$\pm$1.4 & 9.3$\pm$2.2 \\
$N_{\text{SS}}^{\mu,t\bar{t}}$  & 1.0$\pm$0.2 & 2.0$\pm$0.5 & 5.1$\pm$1.2 & 8.1$\pm$1.9 \\
$N_{\text{OS}}^{\mu,t\bar{b}}$  & 0.7$\pm$0.2 & 1.4$\pm$0.4 & 2.1$\pm$0.6 & 4.2$\pm$1.2 \\
$N_{\text{SS}}^{\mu,t\bar{b}}$  & 0.5$\pm$0.1 & 0.8$\pm$0.1 & 1.8$\pm$0.5 & 3.1$\pm$0.9 \\
$f_c^{\mu}$ & 1.195$\pm$0.025$\pm$0.014 & 1.174$\pm$0.027$\pm$0.013 & 1.121$\pm$0.035$\pm$0.013 & 1.148$\pm$0.007$\pm$0.013 \\
$\epsilon_{c}^{\mu}$ & 0.0110$\pm$0.0011$^{+0.0016}_{-0.0017}$ & 0.0122$\pm$0.0013$^{+0.0018}_{-0.0019}$ & 0.0148$\pm$0.0018$^{+0.0022}_{-0.0023}$ & 0.0122$\pm$0.0012$^{+0.0018}_{-0.0019}$ \\
$K_{T}^{\mu}$ & 1.18$\pm$0.02$\pm$0.12 & 1.18$\pm$0.02$\pm$0.12 & 1.18$\pm$0.02$\pm$0.12 & 1.18$\pm$0.02$\pm$0.12 \\
$\frac{\sigma[W(\rightarrow \mu \nu)+c\text{-jet}]}{\sigma[W(\rightarrow \mu \nu)+\text{jets}]}$ & 0.123$\pm$0.037$^{+0.024}_{-0.033}$ & 0.076$\pm$0.050$^{+0.016}_{-0.013}$ & 0.000$\pm$0.058$^{+0.014}_{-0.008}$ & 0.075$\pm$0.031$^{+0.015}_{-0.017}$ \\
\hline
\multicolumn{5}{c}{Combined $W \rightarrow e\nu$ and $W \rightarrow \mu \nu$ decay channels} \\
\hline
$\frac{\sigma[W(\rightarrow \ell \nu)+c\text{-jet}]}{\sigma[W(\rightarrow \ell \nu)+\text{jets}]}$ & 0.097$\pm$0.024$^{+0.016}_{-0.026}$ & 0.091$\pm$0.031$^{+0.016}_{-0.015}$ & 0.025$\pm$0.038$^{+0.005}_{-0.004}$ & 0.074$\pm$0.019$^{+0.012}_{-0.014}$ \\
\end{tabular}
\end{ruledtabular}
\end{table*}
\indent Extraction of samples of $W$+$c$-jet event candidates from the $W$+jets samples follows from selecting events with a $\mu$-tagged jet. \ This jet must contain a reconstructed muon with $p_{T}>4$ GeV and $|\eta|<2.0$ that lies within a cone of $\Delta$$\mathcal{R}(\mu,\text{jet})<0.5$ with respect to the jet axis, and have JES corrected $p_{T}>20$ GeV before including the muon and neutrino energies. \ The muon must be detected in both of the outer two layers of the muon spectrometer, and its muon spectrometer track must be matched to a reconstructed track in the central tracker. \ Background suppression of $Z(\rightarrow\mu^{+}\mu^{-})$+jets events entails rejecting events in which the dimuon invariant mass exceeds $70$~GeV in the muon channel without restricting the charges of the muons. \ Application of all selection criteria yields $N_{\text{OS}}^{e}$ and $N_{\text{SS}}^{e}$ events in the electron channel, and $N_{\text{OS}}^{\mu}$ and $N_{\text{SS}}^{\mu}$ events in the muon channel as reported in Table~\ref{tab:eventY}. \ Estimated multi-jet backgrounds in the $\mu$-tagged jet data samples determined following the matrix method are $N_{\text{OS}}^{e,\text{MJ}}$ and $N_{\text{SS}}^{e,\text{MJ}}$ events in the electron channel, and $N_{\text{OS}}^{\mu,\text{MJ}}$ and $N_{\text{SS}}^{\mu,\text{MJ}}$ events in the muon channel as listed in Table~\ref{tab:eventY}. \ Estimates of the $WW$, $t\bar{t}$ and single top backgrounds from MC are denoted $N_{\text{OS}}^{e,WW}$, $N_{\text{SS}}^{e,WW}$, $N_{\text{OS}}^{e,t\bar{t}}$, $N_{\text{SS}}^{e,t\bar{t}}$, $N_{\text{OS}}^{e,t\bar{b}}$ and $N_{\text{SS}}^{e,t\bar{b}}$ events, respectively, in the electron channel, and $N_{\text{OS}}^{\mu,WW}$, $N_{\text{SS}}^{\mu,WW}$, $N_{\text{OS}}^{\mu,t\bar{t}}$, $N_{\text{SS}}^{\mu,t\bar{t}}$, $N_{\text{OS}}^{\mu,t\bar{b}}$ and $N_{\text{SS}}^{\mu,t\bar{b}}$ events, respectively, in the muon channel as given in Table~\ref{tab:eventY}. \ The estimate of the single top background follows from the $t\bar{b}$ and $t\bar{b}q$ events produced with the \textsc{comphep}~\cite{comphep} generator followed by full detector simulation. \ The quoted uncertainties on the $WW$, $t\bar{t}$ and single top background predictions given in Table~\ref{tab:eventY} are dominated by the uncertainties in their cross section measurements~\cite{WWcross,ttcross,tbcross}. \ Lepton charges are well measured at D0, and uncertainties from charge mis-identification are very small.
\newline\indent The acceptance times efficiencies, $\epsilon_{c}^{\ell}\left(  \ell=e,\mu\right)$, of the $W$+$c$-jet selections relative to inclusive $W$+jets in each $W$ boson decay channel is estimated from the MC simulation, and includes MC to data correction factors estimated using independent data calibration samples. \ The absolute efficiency of reconstructing a $W$ boson with at least one jet cancels in the ratio. \ The relative acceptance includes effects of charm quark to hadron fragmentation, charmed hadron semi-muonic decay and the residual missing calorimeter energy from the muon and neutrino in the $\mu$-tagged jet. \ The relative efficiency accounts for muon identification and track reconstruction effects. \ Charm quark fragmentation and charm hadron decay uncertainties are constrained by previous experiments~\cite{cleocquark,pdg} and contribute $4.5\%$ and $9.5\%$, respectively, to the acceptance uncertainties in both channels. \ The correction included in the acceptance for the missing contribution to the jet $p_T$ from the muon and neutrino energies adjusts the jet $p_T$ spectrum of $W$+$c$-jet candidate events appropriately to the particle level, as verified by a MC closure test. \ A large sample of $J/\psi\rightarrow\mu^{+}\mu^{-}$ events collected at D0 is employed to correct the jet-muon reconstruction efficiency, $(58.7\pm2.8)\%$, computed from the MC simulation, by a factor of $0.89\pm0.06$. \ This correction is found to be independent of the jet $p_{T}$. \ The final acceptance times efficiencies are found to be $\epsilon_{c}^{e}=(1.24\pm0.22)\%$ and $\epsilon_{c}^{\mu}=(1.22\pm0.23)\%$.
\newline\indent The presence of two muons in the muon channel increases the trigger selection efficiency of the $W$+$c$-jet candidates compared to the inclusive $W$+jets data sample. \ The divisor factor $K_{T}^{\mu} = 1.18 \pm 0.12$, extracted from the probability of events being selected when only the jet-muon fires the trigger, corrects for the bias in $W$+$c$-jet selection in the muon channel. \ In the electron channel the factor $K_{T}^{e}$ is unity as the trigger efficiency cancels in the ratio. 
\newline\indent The $W$+$c$-jet cross section ratio is extracted using
\begin{widetext}
\begin{eqnarray*}
\frac{\sigma\left[  W\left(  \rightarrow\ell\nu\right)  +c\text{-jet}\right]
}{\sigma\left[  W\left(  \rightarrow\ell\nu\right)  +\text{jets}\right]  }
 = \frac{\frac{1}{\epsilon_{c}^{\ell} K_{T}^{\ell}}\left[  N_{\text{OS}}^{\ell}-f_{c}^{\ell}\left(N_{\text{SS}}^{\ell}-N_{\text{SS}}^{\ell,\text{MJ}}-N_{\text{SS}}^{\ell,WW}-N_{\text{SS}}^{\ell,t\bar{t}}-N_{\text{SS}}^{\ell,t\bar{b}}\right)-N_{\text{OS}}^{\ell,\text{MJ}}-N_{\text{OS}}^{\ell,WW}-N_{\text{OS}}^{\ell,t\bar{t}}-N_{\text{OS}}^{\ell,t\bar{b}}\right]}{(1-f_{Z}^{\ell}-f_{\text{MJ}}^{\ell})N_{Wj}^{\ell}},
\end{eqnarray*}
\end{widetext}
which requires one further correction in each channel, $f_{c}^{\ell}$, for the
small correlation between the jet-muon and $W$ boson charges that arises in
$W$+light-quark jet events. \ The factor $f_{c}^{\ell}$ is determined from
fully simulated $W$+jet events as the ratio of the predicted number of OS
$\mu$-tagged jets to SS $\mu$-tagged jets in all background samples that pass
the same selection criteria as defined for the data sample. \ Processes
considered include $W$+$u$,$d$,$s$, $W$+$g$, $W$+$c\bar{c}$, $W$+$b\bar{b}$,
and $W$+$c$-jet, where the $c$ quark does not decay semi-muonically in the
last case. \ The $f_{c}^{\ell}$ are parameterized in terms of
jet $p_{T}$ as $f_{c}^{\ell}=a_{\ell}+b_{\ell}\times p_{T}$, with
$a_{e}=1.223\pm0.016$, $a_{\mu}=1.241\pm0.023$, $b_{e}=-0.0017\pm0.0003$, and
$b_{\mu}=-0.0019\pm0.0004$, where all quoted uncertainties of the parameters
are statistical; \ $f_{c}^{\ell}$ decreases with increasing jet $p_{T}$ because the sub-process $q\bar{q}\rightarrow Wg$ dominates $qg\rightarrow
Wq^{\prime}$ at high jet $p_{T}$. \ Systematic uncertainties in $f_{c}^{\ell}$
arise mainly from the cross section and jet fragmentation models. \ The $f_{c}^{\ell}$ are nearly independent of the absolute charged multiplicity per jet and the $W$+light-jets cross section. \ This has been verified by comparing the ratio of all OS tracks to all SS tracks found in jets in the inclusive $W$+jets data sample. \ The $f_{c}^{\ell}$ depend instead on the $K^{\pm}/\pi^{\pm}$ ratio per jet and the relative cross section for $W$ boson plus heavy quark jet final states compared to $W$+light-jets. \ A $6\%$ uncertainty is
assigned to the weighted $\pi^{\pm}$ multiplicity based on a comparison of the
difference between tracking efficiency in data and simulation, and a $20\%$
uncertainty on the $K^{\pm}/\pi^{\pm}$ ratio is estimated based on comparing
$K_{S}^{0}$ production in data to MC. \ Uncertainties in \textsc{alpgen} cross sections are estimated to be $50\%$ for $W$+$b\bar{b}$, $W$+$c\bar{c},$ and $W$+$c$-jet, relative to $W$+light-jets~\cite{singletoppaper}. \ A change of the $W$+$c$-jet cross section by $\pm100\%$ does not lead to a significant effect in $f_{c}^{\ell}$. \ The uncertainty due to PDFs on $f_{c}^{\ell}$ is estimated to be $_{-0.64}^{+0.97}\%$. \ Overall systematic uncertainties are found to be $1.5\%$ for $f_{c}^{e}$ and $1.1\%$ for $f_{c}^{\mu}$, with the relative cross section contributions dominant. \ Adding a $0.6\%$ uncertainty in each channel due to MC statistics yields $f_{c}^{e}=1.149\pm0.018$ and $f_{c}^{\mu
}=1.148\pm0.015$ averaged over all $p_{T}>20$ GeV.
\begin{figure}[ptb]
\includegraphics[scale=0.43]{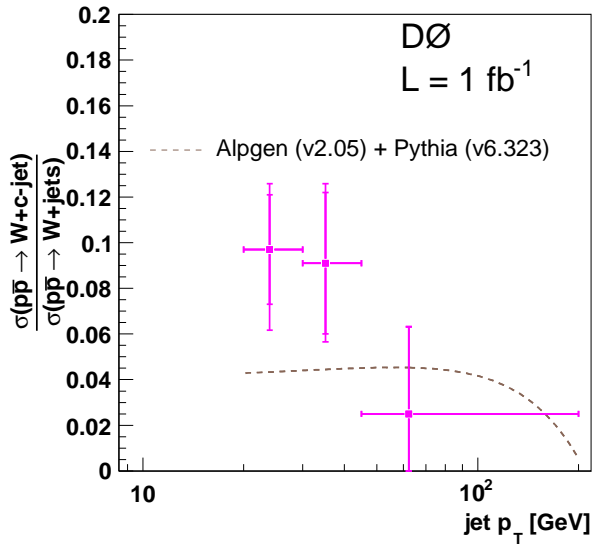}
\caption{Measured ratio $[\sigma(W$+$c$-jet$)/\sigma(W$+jets$)]$ for jet $p_{T}>20$ GeV and $|\eta|<2.5$. \ The inner error bars around the data points show the statistical only uncertainties and the full bars represent the quadratic sum of statistical and systematic uncertainties. \ The systematic uncertainty on $W$+$c$-jet fraction includes the uncertainties due to JES, the jet $p_{T}$ resolution, the background correction factor $f_{c}^{\ell}$, and the product of the relative acceptance and efficiencies $\epsilon_{c}^{\ell}$. \ It also includes the uncertainty due to $K_{T}^{\mu}$ in the muon channel.}
\label{fig:result3binscomb}
\end{figure}
\begin{figure*}[ptb]
\psfrag{sgn [a/sigma(a)] of muon track} {sgn[$\frac{a}{\sigma_a}$] of muon track}
\psfrag{jet M/pT} {jet $\frac{M}{p_{T}}$} 
$\begin{array}
[c]{c@{\hspace{0.01in}}c}
\includegraphics[scale=0.46]{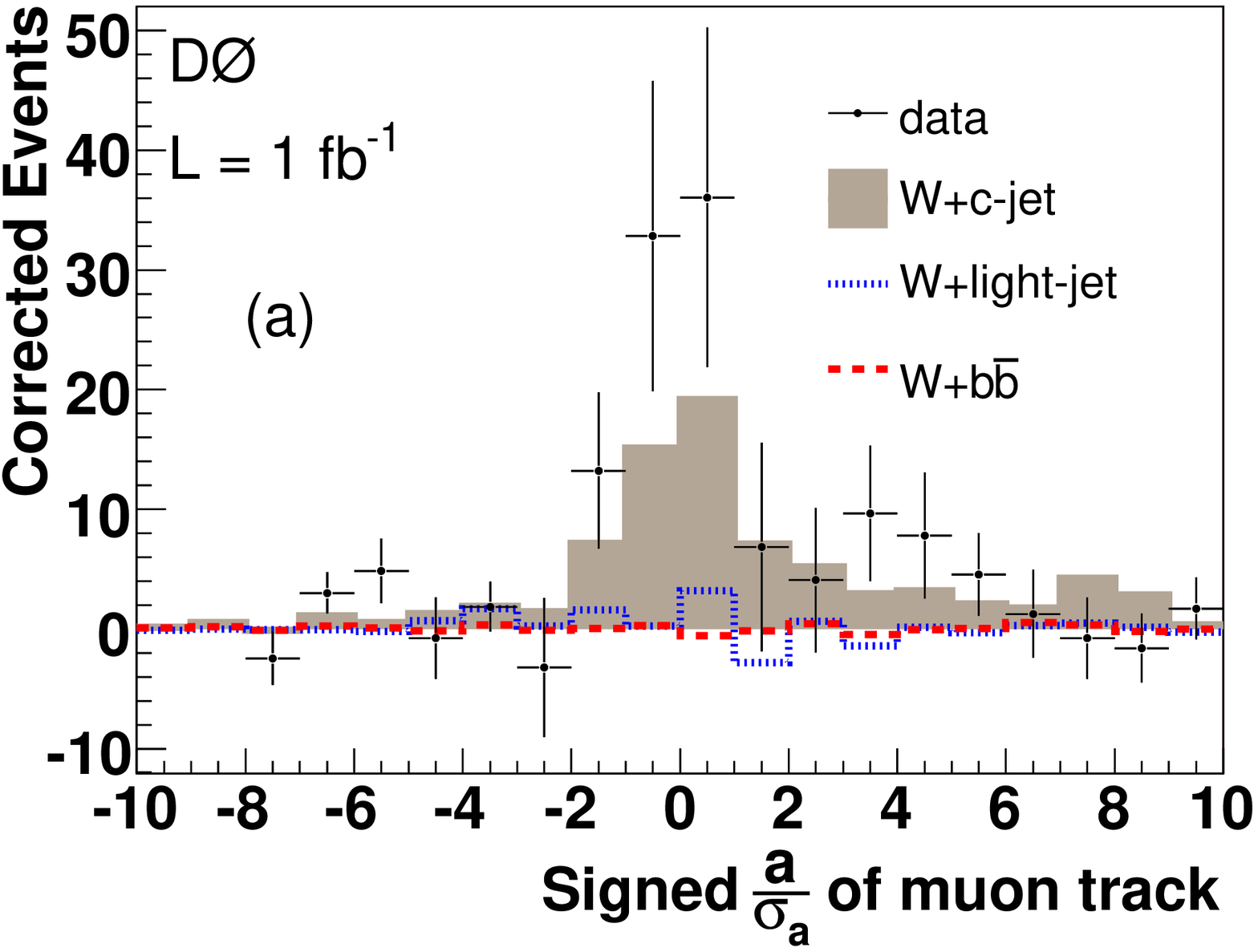} &
\includegraphics[scale=0.46]{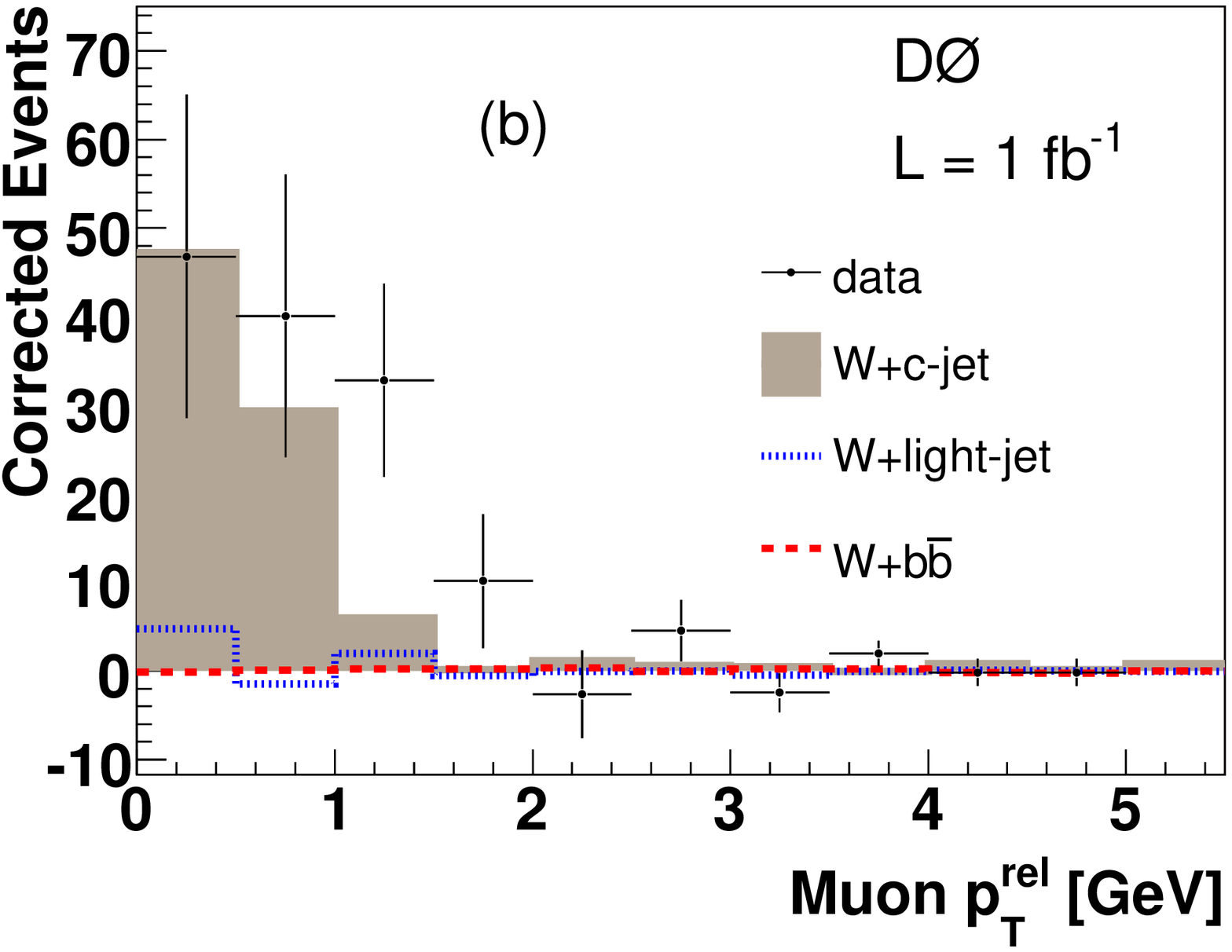}\\
\end{array}$
\caption{Comparison of the background-subtracted ($N_{\text{OS}}^{\ell}-f_{c}^{\ell}N_{\text{SS}}^{\ell}$) data in the combined electron and muon channels with the simulation. \ (a) \textit{{Signed}} significance in impact parameter of the jet-muon track with respect to the interaction point, (b) jet-muon transverse momentum relative to the jet axis ($p_T^{\text{rel}}$).}
\label{fig:ipsig}
\end{figure*}
\newline\indent Table~\ref{tab:eventY} summarizes the cross section ratio
measurements and their uncertainties for the electron and the muon channels
for all jet $p_{T}>20$ GeV and jet $|\eta|<2.5$, and for three jet $p_{T}$
bins with $|\eta|<2.5$ in each bin. \ The ratio measurements benefit from cancellation of several uncertainties, notably the integrated luminosity~\cite{d0lumi}, lepton detection efficiency, and jet energy scale (JES). \ Table~\ref{tab:syst} lists remaining absolute systematic uncertainties on the measurement estimated from the MC simulation. \ These arise mainly from second order JES effects, jet $p_{T}$ resolution (JPR),  $c$-jet tagging efficiency,
and the $W$+$c$-jet background correction factors $f_{c}^{\ell}$.
\newline\indent \ The measured $W$+$c$-jet fractions integrated over
$p_{T}>20$ GeV and $|\eta|< 2.5$ are 
\begin{eqnarray*}
\frac{\sigma\left[W\left(\rightarrow e\nu\right)+c\text{-jet}\right]}{\sigma\left[W\left(\rightarrow e\nu\right)+\text{jets}\right]}&=&0.073\pm0.023 (\text{stat.})_{-0.014}^{+0.012}(\text{syst.}),\\
\frac{\sigma\left[W\left(  \rightarrow\mu\nu\right)  +c\text{-jet}\right]}{\sigma\left[W\left(\rightarrow\mu\nu\right)+\text{jets}\right]}&=&0.075\pm0.031(\text{stat.})_{-0.017}^{+0.015}(\text{syst.}).
\end{eqnarray*} 
\noindent Since the $W\rightarrow e\nu$ and $W\rightarrow\mu\nu$ measurements are consistent with one another, and statistical uncertainties dominate, the two lepton channels are combined to yield 
\begin{eqnarray*}
\frac{\sigma\left[W+c\text{-jet}\right]}{\sigma\left[ W+\text{jets}\right]}=0.074\pm0.019(\text{stat.})_{-0.014}^{+0.012}(\text{syst.}).
\end{eqnarray*}
\noindent Systematic uncertainties are taken to be fully correlated in the two channels. \ This measurement can be compared to  $W$+$c$-jet fraction predicted by \textsc{alpgen} and \textsc{pythia} of 0.044$\pm$0.003, where the quoted theoretical uncertainty derives from the uncertainty on the \textsc{cteq6.5} PDFs. \ Due to the relatively small contributions of $W$+$b\bar{b}$ and $W$+$c\bar{c}$ to inclusive $W$+jets, the model prediction of the $W$+$c$-jet rate has $\lesssim 5\%$ sensitivity to their cross sections. \ Figure~\ref{fig:result3binscomb} shows the differential $W$+$c$-jet fraction, and compares the data to a model prediction using leading order QCD augmented by \textsc{alpgen} and \textsc{pythia}.
\begin{table}[ptb]
\caption{Fractional systematic uncertainties on the measurement in the $W\rightarrow e\nu$ and the $W\rightarrow \mu\nu$ channels.}%
\label{tab:syst}%
\renewcommand{\arraystretch}{1.5}
\centering
\begin{ruledtabular}
\begin{tabular}[c]{c|cccc|ccccc}
& \multicolumn{4}{c}{$e$ channel} & \multicolumn{5}{c}{$\mu$ channel} \\ 
\hline
$p_T$ & JES & JPR & $f_{c}^{e}$ & $\epsilon_c^{e}$ & JES & JPR & $f_{c}^{\mu}$ & $\epsilon_c^{\mu}$ & $K_{T}^{\mu}$ \\
GeV & $\%$ & $\%$ & $\%$ & $\%$ & $\%$ & $\%$ & $\%$ & $\%$ & $\%$ \\
\hline
$20$--$30$  & $_{-21.6}^{+0}$ & $_{-4.8}^{+2.4}$ & $_{-4.1}^{+3.8}$ & $_{-15.6}^{+15.7}$ & $_{-17.6}^{+0}$ & $_{-7.5}^{+5.0}$ & $_{-3.3}^{+2.5}$ & $_{-16.2}^{+15.3}$ & $\pm10$ \\
$30$--$45$  & $_{-4.3}^{+6.4}$ & $_{-4.3}^{+2.1}$ & $_{-4.7}^{+4.3}$ & $_{-14.4}^{+14.5}$ & $_{-0.7}^{+9.8}$ & $_{-6.7}^{+4.5}$ & $_{-4.3}^{+3.1}$ & $_{-14.9}^{+14.1}$ & $\pm10$\\
$45$--$200$ & $_{-2.2}^{+2.2}$ & $_{-4.5}^{+2.2}$ & $_{-7.6}^{+6.9}$ & $_{-14.6}^{+14.7}$ & $_{-0}^{+27.7}$ & $_{-7.0}^{+4.7}$ & $_{-5.2}^{+4.0}$ & $_{-15.8}^{+15.0}$ & $\pm10$ \\
\hline
$20$--$200$  & $_{-9.0}^{+0}$ & $_{-4.5}^{+2.3}$ & $_{-5.2}^{+4.5}$ & $_{-15.0}^{+15.1}$ & $_{-2.4}^{+5.9}$ & $_{-7.1}^{+4.7}$ & $_{-4.3}^{+3.1}$ & $_{-15.7}^{+14.9}$ & $\pm10$ \\
\end{tabular}
\end{ruledtabular}
\end{table}
\newline \indent As a test of the $W$+$c$-jet signal hypothesis, Fig.~\ref{fig:ipsig}(a) compares data to \textsc{alpgen} and \textsc{pythia} expectations in the background-subtracted distribution of the signed impact parameter significance, $a/\sigma_{a}$, for the jet-muon, where $a$ is the projected distance of closest approach of the jet-muon to the event interaction point in the transverse plane, and $\sigma_{a}$ is the estimated uncertainty on $a$. \ Data show satisfactory agreement with expectations for $W$+$c$-jet production and the underlying OS-SS ansatz after the subtraction of light and $b$ quark jet contributions. \ Similarly, the distribution of the relative transverse momentum of the jet-muon with respect to the jet axis, $p_{T}^{\text{rel}}$, shows the consistency between data and the $c$-jet expectation as illustrated in Fig.~\ref{fig:ipsig}(b). 
\newline \indent To quantify the probability that the observed excess of OS events over SS events is due exclusively to background fluctuations, ensembles for OS, SS backgrounds and inclusive $W$+jets are generated that incorporate all the systematic uncertainties together with the correlations among the OS, SS backgrounds and $W$+jets expectations using Gaussian samplings of the uncertainties. \ The probability that background fluctuations could produce the observed fraction of the signal events in the inclusive $W$+jets sample is $2.5\times 10^{-4}$, corresponding to a $3.5$ $\sigma$ significance for the $W$+$c$-jet hypothesis.
\newline \indent In conclusion, we have performed a measurement of the $W$+$c$-jet/$W$+jets cross section ratio at a hadron collider using both electron and muon decay channels of the $W$ boson and utilizing the correlation between the charge of the jet-muon with that of the $W$ boson. \ The probability that background fluctuations could produce an estimated $W$+$c$-jet fraction in $W$+jets equal to or larger than the one measured in data is $2.5 \times 10^{-4}$, which corresponds to a $3.5$ $\sigma$ significance of the observation. \ We find our measurement to be consistent with LO perturbative QCD predictions of the $W$+$c$-jet production rate and with an $s$ quark PDF evolved from $Q^{2}$ scales two orders of magnitude below those of this measurement. \ The measurement further provides direct experimental evidence of the underlying partonic process $qg\rightarrow W q^{\prime}$ that should dominate $W$ boson production at the CERN Large Hadron Collider (LHC).\\ \indent
%
We thank the staffs at Fermilab and collaborating institutions, 
and acknowledge support from the 
DOE and NSF (USA);
CEA and CNRS/IN2P3 (France);
FASI, Rosatom and RFBR (Russia);
CNPq, FAPERJ, FAPESP and FUNDUNESP (Brazil);
DAE and DST (India);
Colciencias (Colombia);
CONACyT (Mexico);
KRF and KOSEF (Korea);
CONICET and UBACyT (Argentina);
FOM (The Netherlands);
STFC (United Kingdom);
MSMT and GACR (Czech Republic);
CRC Program, CFI, NSERC and WestGrid Project (Canada);
BMBF and DFG (Germany);
SFI (Ireland);
The Swedish Research Council (Sweden);
CAS and CNSF (China);
and the
Alexander von Humboldt Foundation.

\end{document}